\documentclass[aps,prb,twocolumn,superscriptaddress,floatfix,showpacs,amsmath,amssymb,nofootinbib,longbibliography,nobibnotes,noeprint]{revtex4-2}

\usepackage{graphicx}
\usepackage{dcolumn,xcolor}
\usepackage{bm}
\usepackage{color}
\usepackage{tabularx}
\usepackage{footmisc}
\usepackage{siunitx}
\DeclareSIUnit{\formulaunit}{\text{f.u.}}
\DeclareSIUnit{\erg}{\text{erg}}
\DeclareSIUnit{\gauss}{\text{G}}
\usepackage{nicefrac}
\usepackage{amsmath}
\usepackage{verbatim}
\usepackage{parskip,stackengine}
\usepackage{soul,xr}
\usepackage[version=3]{mhchem}
\usepackage[normalem]{ulem}

\usepackage{xr} 
\newcommand{\fgt}{Fe\(_3\)GeTe\(_2\)}

\newcommand{\cri}{CrI\(_3\)}
\newcommand{\crbr}{CrBr\(_3\)}
\newcommand{\cst}{CrSiTe\(_3\)}
\newcommand{\cgt}{CrGeTe\(_3\)}
\newcommand{\etal}{\textit{et~al.}}

\DeclareSIUnit{\mb}{$\mu_{\mathrm{B}}$}
\DeclareSIUnit{\fu}{\mathrm{f.u.}}

\newcommand{\tc}{$T_{\mathrm{C}}$}

\begin{document}
	
\title{Elucidating the origin of long-range ferromagnetic order in Fe$_3$GeTe$_2$ by low-energy magnon excitation studies}
	
	
\author{B.~Beier}
\affiliation{Kirchhoff Institute for Physics, Heidelberg University, INF 227, D-69120 Heidelberg, Germany}
	
\author{E.~Walendy}
\affiliation{Kirchhoff Institute for Physics, Heidelberg University, INF 227, D-69120 Heidelberg, Germany}

\author{J.~Arneth}
\affiliation{Kirchhoff Institute for Physics, Heidelberg University, INF 227, D-69120 Heidelberg, Germany}
	
\author{E.~Brücher}
\affiliation{Max Planck Insitute for Solid State Research, Heisenbergstraße 1, D-70569 Stuttgart, Germany}

\author{R. K.~Kremer}
\affiliation{Max Planck Insitute for Solid State Research, Heisenbergstraße 1, D-70569 Stuttgart, Germany}

\author{R.~Klingeler}\email{klingeler@kip.uni-heidelberg.de}
\affiliation{Kirchhoff Institute for Physics, Heidelberg University, INF 227, D-69120 Heidelberg, Germany}

\date{\today}
	
\begin{abstract}
We report a detailed high-field/ high-frequency ferromagnetic resonance (HF-FMR) study of low-energy magnon excitations in the van der Waals ferromagnet Fe$_3$GeTe$_2$. At $\SI{2}{\kelvin}$, the field dependence of the magnon branches is well described by a semiclassical domain-based model, from which we extract key microscopic parameters including the anisotropy gap $\Delta = \SI{170\pm4}{\giga\hertz}$, the anisotropy field $B_{\rm A} = \SI{5.85\pm0.08}{\tesla}$, and the effective $g$-factor $g_{\rm ab}\simeq g_{\rm c} = 2.07(4)$. Furthermore the uniaxial anisotropy constant was determined to be $K = \num{10.5\pm0.23}\times \SI{e6}{\erg\per\cubic\centi\meter}$. Anisotropic short-range magnetic order persists above \tc~ up to approximately $\SI{270}{\kelvin}$, as evidenced by a finite anisotropy gap and anisotropic shifts in the FMR resonance fields. Both results clearly show the presence of anisotropic local magnetic fields well above \tc . Our findings underscore the crucial role of magneto-crystalline anisotropy in driving long-range magnetic order in Fe$_3$GeTe$_2$.
\end{abstract}
	
\maketitle
    
\section{Introduction}

Reducing one dimension of a material below characteristic length scales to achieve two-dimensional (2D) systems can yield spectacular novel properties as evidenced among many other examples by the Quantum Hall effect, Dirac-fermions in graphen or the appearance of topological phases
~\cite{Klitzing1980,Novoselov2005,Ren2016,Tsui1982}. The pronounced strength of thermal as well as of quantum fluctuations in systems with reduced dimensionality in particular challenges long-range (magnetic) ordering phenomena of the corresponding bulk materials~\cite{Mermin1966,Han2012,Kasahara2018,Xu2020a} and hence opens a wide field for fundamental research but also offers new routes for technological applications~\cite{Wang2018,Kim2018}. The recent advances in the research field of 2D van-der-Waals (vdW) materials illustrate these prospects as progress in our understanding of low-dimensional materials is accompanied by exploiting these properties in actual devices (see, e.g.,~\cite{Gong2019Review,Novoselov2016,Liu2016} and references therein). In this respect, the existence of long-range ferromagnetic (FM) order is particularly important for magnetic devices. Since in the strict 2D Heisenberg case, long-range magnetic order does not evolve for any short-range in-plane magnetic exchange~\cite{Mermin1966}, magnetic anisotropy is a crucial ingredient for establishing ferromagnetism and to drive it beyond room temperature~\cite{Alahmed2021}. While several semiconducting FM 2D vdW  materials are known (e.g., \cri\ \cgt\ , \cst\ , \crbr\ , see~\cite{Huang2017, Gong2017, Ouvrard1988, Ichiro1960} and for a review~\cite{Kataev2024}), the family of Fe\(_n\)GeTe\(_2\) ($n\simeq3,4,5$) materials offers the rare case of metallic 2D ferromagnets~\cite{deiseroth2006,Seo2020,May2019,May2019} with FM ordering temperatures \tc\ approaching or even exceeding room temperature~\cite{Khan2019,Alahmed2021} (see also Table~\ref{tab:list}). From this family, \fgt\ offers great potential for application in magnetic heterostructures and spintronics which is further boosted by its good air stability~\cite{deiseroth2006}. Recent studies exploiting the low-dimensional metallic nature of \fgt\ have found, e.g., large anomalous Hall effect~\cite{Kim2018}, evidence of Kondo lattice physics and heavy-fermion states~\cite{Zhang2018}, skyrmions~\cite{Ding2020} and a moderate magneto-caloric effect~\cite{Verchenko2015}. 

\begin{figure}[htb]
	\centering
	\includegraphics[width=\columnwidth,clip]{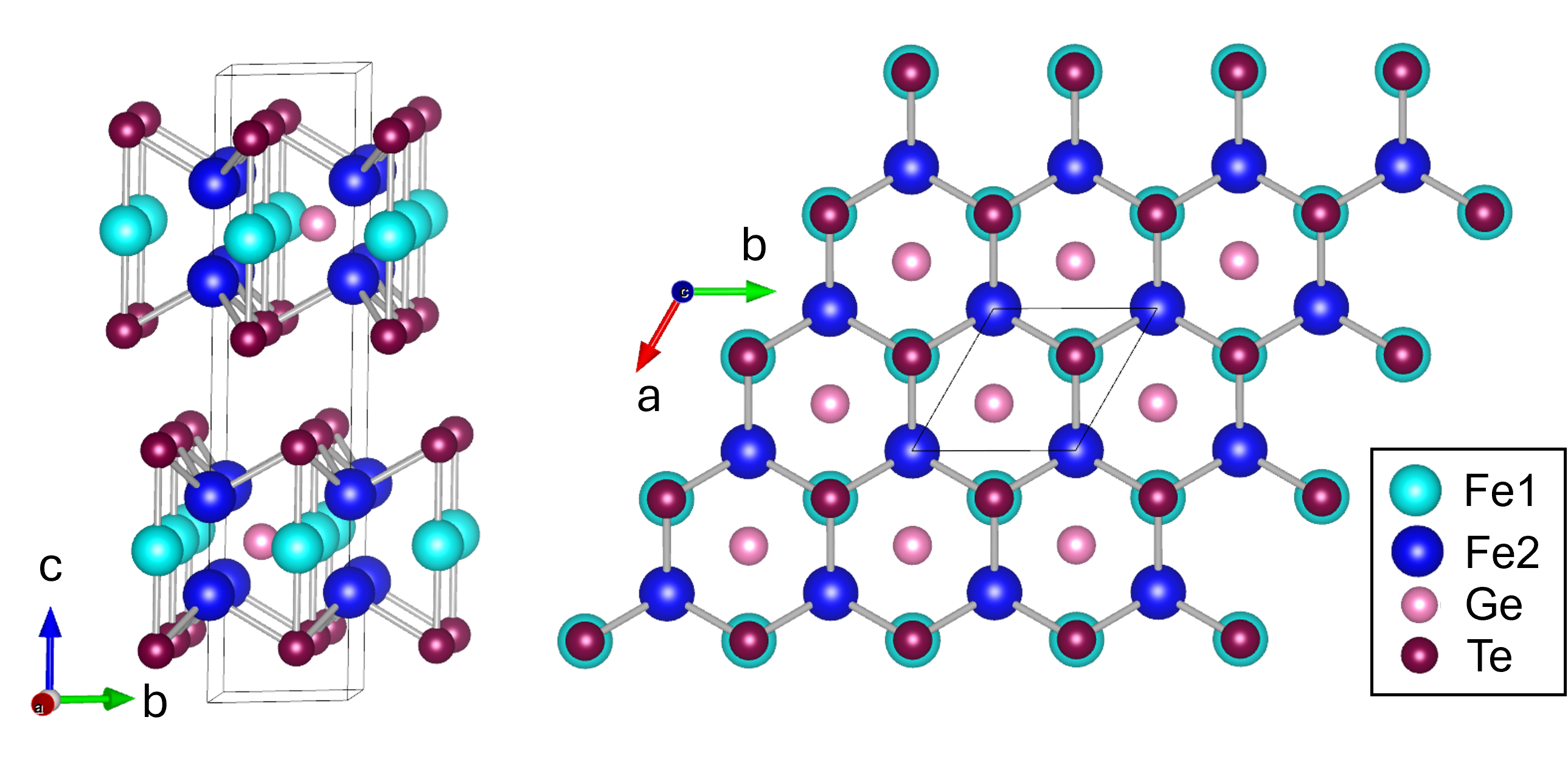}
	\caption{Crystal structure of \fgt~\cite{deiseroth2006}: (a) Unit cell and (b) top view of $ab$ plane. The figures were generated with VESTA~\cite{vesta}.}
	\label{fig:crystal_structure}
\end{figure}

\fgt\ features iron germanium layers which are sandwiched by tellurium atoms and only weakly bound via vdW bonds~\cite{deiseroth2006} (see Fig.~\ref{fig:crystal_structure}). The projection of the iron atoms to the $ab$ plane shows a hexagonal arrangement with germanium atoms in the center of each hexagon as shown in Fig.~\ref{fig:crystal_structure}b. There are two differently coordinated iron sites Fe1 and Fe2 which are positioned at different heights along the $c$ axis in each hexagon. Even for samples prepared under stoichiometric conditions, the Fe2 site is not fully occupied~\cite{kremer2024}. 

In comparison with other FM vdW materials, \fgt~exhibits high uniaxial magnetocrystalline anisotropy (see Table~\ref{tab:list}) and its magnetic behavior can be tuned through variation of iron deficiency~\cite{Liu2017b}, the number of layers~\cite{Fei2018}, or by strain tuning~\cite{Hu2020} (see also the Review~\cite{Papavasileiou2023}). Despite extensive previous research, the microscopic origin of long-range FM order in \fgt\ still remains an open question. Prior studies focusing on whether the FM ordering arises from itinerant or localized magnetic moments lead to contradictory conclusions~\cite{Chen2013,Zhang2018,Xu2020b}.

In this work, we investigate low-energy magnon excitations in \fgt\ using high-field/high-frequency ferromagnetic resonance (HF-FMR). The field dependence of the magnon branches at \SI{2}{\kelvin} yields the size of the anisotropy gap, the anisotropy field, and the effective $g$-factor of \fgt\ which are also used to determine the uniaxial anisotropy constant. Furthermore, from HF-FMR measurements at various temperatures we find that long-range order develops from anisotropic short-range order which is quasi-static on the GHZ-time-scale in a wide temperature regime above \tc . Our findings underscore the crucial role of magneto-crystalline anisotropy and elucidate the origin of long-range magnetic order in Fe$_3$GeTe$_2$.
	
\section{Experimental}
	
High-field/ high-frequency electron spin resonance (HF-ESR) measurements were performed in the frequency range 40~GHz $\leq f \leq$ 850~GHz in transmission mode and Faraday configuration. The generation and detection of the microwave radiation was facilitated by means of a millimetre-wave vector network analyzer from AB Millimetre~\cite{comba2015}. The measurements were performed in a magnetocryostat system (Oxford) equipped with a 16~T superconducting coil and a VTI temperature control insert operating in the range 1.7~K~$\leq T \leq$~300~K~\cite{werner2017}. The measurements have been performed on an approximately rectangularly-shaped thin single crystal of dimensions $2.8\times 2.7 \times 0.3$~mm$^3$ fixed in a brass ring (see also the Supplemental Materials (SM)~\cite{SM}). The sample originates from the same batch as in~\cite{kremer2024}, where the atomic site occupancy was determined to be Fe\(_{2.92(1)}\)Fe\(_{1.02(3)}\)Te\(_2\)~\cite{kremer2024}, implying an identical composition for our sample. The static magnetization was measured in a Quantum Design MPMS3 superconducting quantum interference device (SQUID) magnetometer operating with the magnetic field applied parallel to the crystallographic $c$ axis of the material. Our data (see $M(T,B=1~{\rm T})$ in Fig.~\ref{fig:magnetization} in the SM~\cite{SM}) show identical behavior as reported in~\cite{kremer2024} where from modified Arrot-Belov plots \tc~=~\SI{217}{\kelvin} was determined.
	
\section{Results}
		
\subsection{Ferromagnetic resonance modes at $T=2$~K}

\begin{figure*}[htb]
	\centering
	\includegraphics[width=\textwidth,clip]{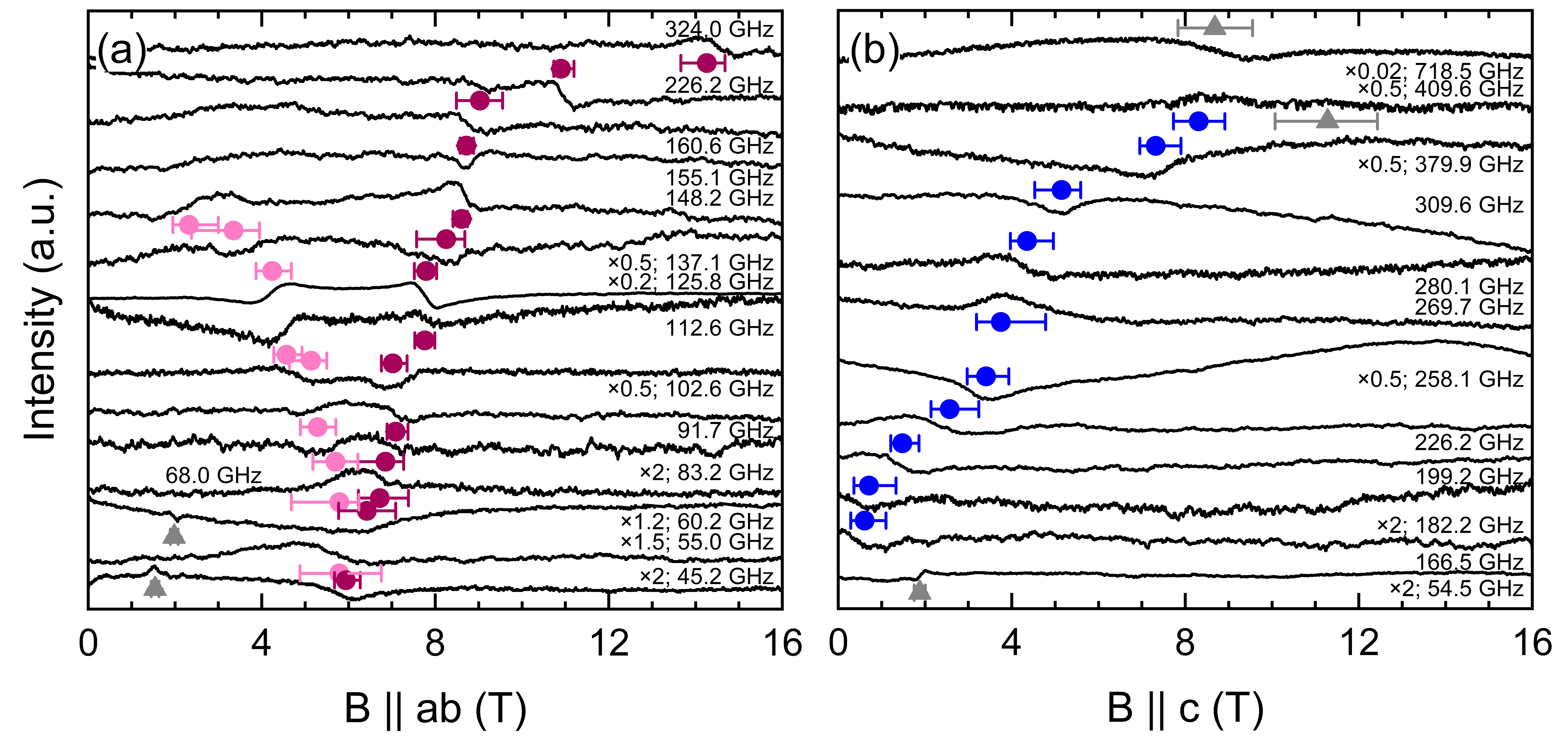}
	\caption{HF-ESR spectra at selected frequencies obtained at $T=2$~K for (a) $B||ab$ plane ($\nu_1$, $\nu_2$) and (b) $B||c$ axis ($\nu_3$). Colored symbols mark the positions of the resonance features. Some spectra were corrected for a linear or quadratic background.}
	\label{fig:spectra_2K}
\end{figure*}

\begin{figure}[htb]
\centering
\includegraphics[width=\columnwidth,clip]{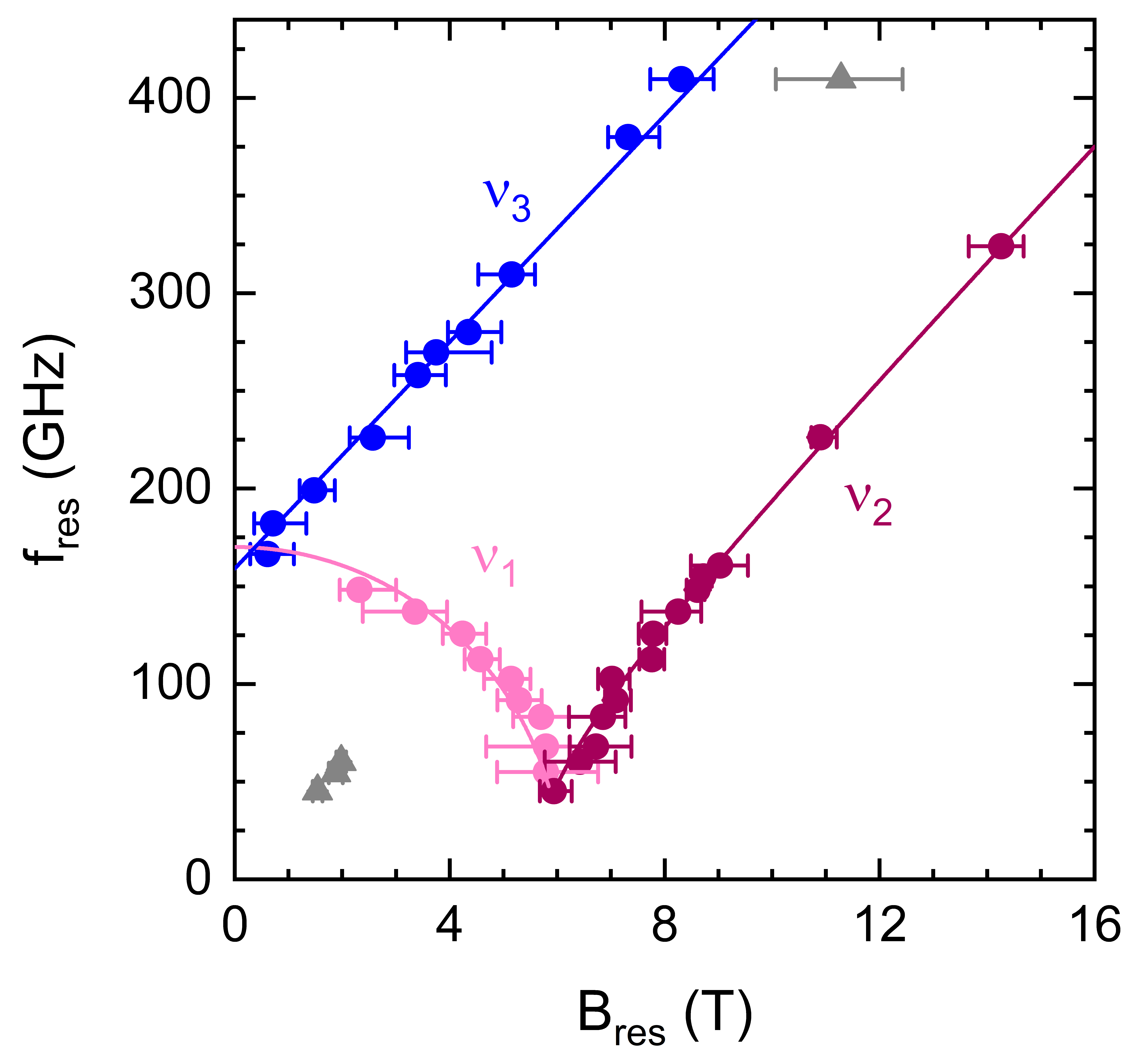}
\caption{Resonance-frequency–magnetic-field diagram of \fgt\ at \SI{2}{\kelvin}. The resonance branches for $B||ab$ plane ($B||c$) are labeled $\nu_{\rm 1}$/$\nu_{\rm 2}$ ($\nu_3$). 
Solid lines are fits to the data using the domain-based model described in the text.}
\label{fig:fb-diagram_2K}
\end{figure}

Firstly, we report the magnetic field dependence of the ferromagnetic magnon modes in \fgt\ at low temperature. The FMR spectra were obtained at $T = \SI{2}{\kelvin}$, i.e., well below the ferromagnetic ordering temperature, and show several resonance features associated with magnon excitations. The corresponding spectra for $B||c$ axis and $B||ab$ plane are shown in Fig.~\ref{fig:spectra_2K}. The resonance features are marked by colored symbols and can be clearly identified in the spectra as the amplitude and the phase of the signal was measured in both up- and down-sweeps of the field. For $B || c$ axis a single resonance branch $\nu_3$ is detected. In contrast, there are two distinct resonance features $\nu_1$ and $\nu_2$ in the spectra obtained for $B || ab$ plane, the former being observed only up to $f = \SI{148.2}{\giga \hertz}$. There are a few further resonance features (grey symbols in Figs.~\ref{fig:spectra_2K} and \ref{fig:fb-diagram_2K}) which stem from the kapton sealing tape (e.g., at $\SI{2}{\tesla}$ and $f = \SI{54.5}{\giga \hertz}$) 
or cannot be attributed to any distinct resonance branch. 
The corresponding resonance fields are summarized in the resonance-frequency-magnetic-field diagram ($fB$ diagram) in Fig.~\ref{fig:fb-diagram_2K}. The $fB$ diagram shows the three distinct resonance branches mentioned above which main qualitative features are as following: (1) Two modes ($\nu_1$, $\nu_3$) demonstrate a zero-field excitation gap of about $\tilde \Delta \simeq 170$~GHz. (2) The in-plane modes $\nu_1$, $\nu_2$ soften at about $\simeq 6$~T with distinct ($\nu_1$) or modest ($\nu_2$) bending. (3) The high-frequency behaviors 
of $\nu_2$ and $\nu_3$ are quasi-linear with approximately the same slopes. 

To model the field dependence of the three distinct resonance branches in detail, the domain-based model for FMR modes in materials with the easy-axis parallel to the crystallographic $c$ axis and in Faraday-configuration is applied. The model predicts three FMR modes -- i.e., the solutions of the Larmor equations obtained by using the Smit-Beljers approach -- with the following field-dependencies of the resonance frequencies~ \cite{Li2021,Smit1955}:

\begin{widetext} 
\begin{equation}
		\left( \frac{\nu_{\rm1}}{\gamma_{\rm ab}} \right)^2 = \left(B_{\rm A}+N_{\rm x}M_{\rm s}\right)\left(B_{\rm A}+M_{\rm s}\sin(\alpha)^2\right)
		-\frac{\left(B_{\rm A}+M_{\rm s}\sin(\alpha)^2-N_{\rm z}M_{\rm s}\right)\left(B_{\rm A}+N_{\rm x}M_{\rm s}\right)}{\left(B_{\rm A}+N_{\rm y}M_{\rm s}\right)^2}B_{\rm{res}}^2 \text{~(for~} B_{\rm {res}}||ab\leq B_{\rm {sat}}\text{)}
		\label{eq:FMR_nu1_domain-based}
    \end{equation}

	\begin{equation}
		\begin{split}
			\left( \frac{\nu_{\rm 2}}{\gamma_{\rm {ab}}} \right)^2 = \left(B_{\rm {res}}-\left(B_{\rm A}-\left(N_{\rm z}-N_{\rm y}\right)M_{\rm s}\right)\right)			\times \left(B_{\rm {res}}-\left(N_{\rm y}-N_{\rm x}\right)M_{\rm s}\right) \text{~(for~} B_{\rm {res}}||ab\geq B_{\rm {sat}}\text{)}
			\label{eq:FMR_nu2_domain-based}
		\end{split}
	\end{equation}
\end{widetext}		
	\begin{equation}
		\begin{split}
			\frac{\nu_{\rm3 }}{\gamma_{\rm c}} = B_{\rm {res}}+B_{\rm{A}}-N_{\rm z}M_{\rm s} \text{~(for~} B_{\rm {res}}||c\text{)}.
			\label{eq:FMR_nu3_domain-based}
		\end{split}
	\end{equation}
		
Here, $\alpha$ corresponds to the angle a certain magnetic domain wall encloses with the external applied magnetic field, $B_{\rm A}$ is the anisotropy field, $M_{\rm s}$ the saturation magnetization, $N_{\rm x}$, $N_{\rm y}$ and $N_{\rm z}$ are the demagnetization factors, and $\gamma_{\rm{ab/c}}$ are the gyromagnetic ratios for $B || ab$/$B||c$ in units of \SI{}{\giga \hertz \per \tesla}. The gyromagnetic ratio is then given by $(\nicefrac{\mu_{\rm B}}{h})\times g_{\rm {ab/c}}$, where $g_{\rm{ab/c}}$ are the effective $g$-factors, related to the slope of the resonance branches in the $fB$-diagram for the respective field directions. At high magnetic fields $B\geq B_{\rm sat} = B_{\rm A}+N_{\rm y}M_{\rm s}$, the system adopts the fully polarized spin configuration which is well described by a single domain model and the respective resonance branch $\nu_{\rm 2}$ (Eq.~\ref{eq:FMR_nu2_domain-based}).
			
The demagnetization factors were determined to $N_{\rm x} = \num{0.097 \pm 0.004}$, $N_{\rm y} = \num{0.099\pm0.005}$, $N_{\rm z} = \num{0.804\pm0.009}$ by assuming a perfect rectangularly-shaped crystal with the above-mentioned dimensions using the formulas from Ref.~\cite{Aharoni1998}. To account for potential deviations from these values, e.g., due to imperfect rectangular shape or the finite penetration depth in the metallic sample, fits of the domain-based model were also performed with the demagnetization factors 
as free parameters. The determined values for $B_{\rm A}$, $\alpha$ and $\gamma_{\rm{ab/c}}$ did not change significantly. Therefore to determine $B_{\rm A}$, $\alpha$ and $\gamma_{\rm {ab/c}}$ from the experimental data, the above-mentioned values for $N_{\rm x}$-$N_{\rm z}$ calculated for a perfect rectangularly-shaped crystal were used as an approximation for the demagnetization factors of the non-perfectly-shaped crystal measured in this work. The saturation magnetization was determined from the isothermal magnetization data (see Fig.~\ref{fig:magnetization} in the SM~\cite{SM}) to $M_{\rm S}=\SI{4.36\pm0.07}{\mu_{\rm B} \per \fu}$. 	

As illustrated in Fig.~\ref{fig:fb-diagram_2K}, the observed resonance branches $\nu_{\rm{1}}$-$\nu_{\rm 3}$ are well described by Equations~\ref{eq:FMR_nu1_domain-based}-\ref{eq:FMR_nu3_domain-based}. The experimental data of all branches were simultaneously fitted. The three lowest lying resonance features ($f = \SI{45.2}{\giga \hertz}$, $\SI{55.0}{\giga \hertz}$, $\SI{60.2}{\giga \hertz}$) were assigned to the resonance branches $\nu_{\rm{1}}$ and $\nu_{\rm 2}$ such that the deviation of the fit from the data is minimized. The best fit to the data yields $B_{\rm A} = \SI{5.85\pm0.08}{\tesla}$, $g_{\rm{ab}} = g_{\rm c} = \num{2.07\pm0.04}$ and $\alpha = \SI{0\pm11}{\degree}$.

The domain-based model applied here assumes the presence of magnetic domains with different orientations of the magnetization within neighboring domains. For comparison, we have fitted the field dependence of the three magnon branches at \SI{2}{\kelvin} also by a single domain model where $\tilde{\nu_{\rm 1}}$ is described by Eq.~\ref{eq:FMR_nu1_single-domain} in the SM~\cite{SM}. This single-domain model however yields an unsatisfactory description of the experimental data. In particular it fails to describe the resonance branch $\nu_1$ since it does not capture the experimentally observed pronounced bending which is reproduced well in the domain-based model (see Fig.~\ref{fig:fb-diagram_2K_supp} in the SM~\cite{SM}). Our experimental data hence reveals the presence of domains in \fgt.

The experimental data presented in Fig.~\ref{fig:fb-diagram_2K} clearly show the presence of a zero-field excitation gap $\tilde\Delta$ which is attributed to magnetic anisotropy. While its approximate value can be directly read-off in the $fB$-diagram, its more precise determination is feasible by exploiting the fit results of the domain-based model to the resonance branches $\nu_{\rm 1}$-$\nu_{\rm 3}$. This procedure yields $\Delta = \gamma B_{\rm A} = \SI{170\pm4}{\giga \hertz}$ for the anisotropy gap at $T=\SI{2}{\kelvin}$. $\Delta$ is associated with the $f$-axis intersection (i.e., $\tilde \Delta$) of the resonance branches $\nu_1$ and $\nu_3$ of the domain-based model. Calculating this intersection, e.g., for $\nu_3$ by setting $B_{\rm{res}} = 0$ in eq.~\ref{eq:FMR_nu3_domain-based} yields $\tilde \Delta = \gamma_{\rm c} B_{\rm A} - \gamma_{\rm c} N_{\rm z} M_{\rm S}$ and therefore the relation $\Delta = \gamma_{\rm c} B_{\rm A} = \tilde{\Delta}+\gamma_c N_z M_S$.

\subsection{Temperature dependence of the FMR modes and quasi-static magnetic fields well above $T_{\rm C}$}

\begin{figure}[htb]
\centering
\includegraphics[width=\columnwidth,clip]{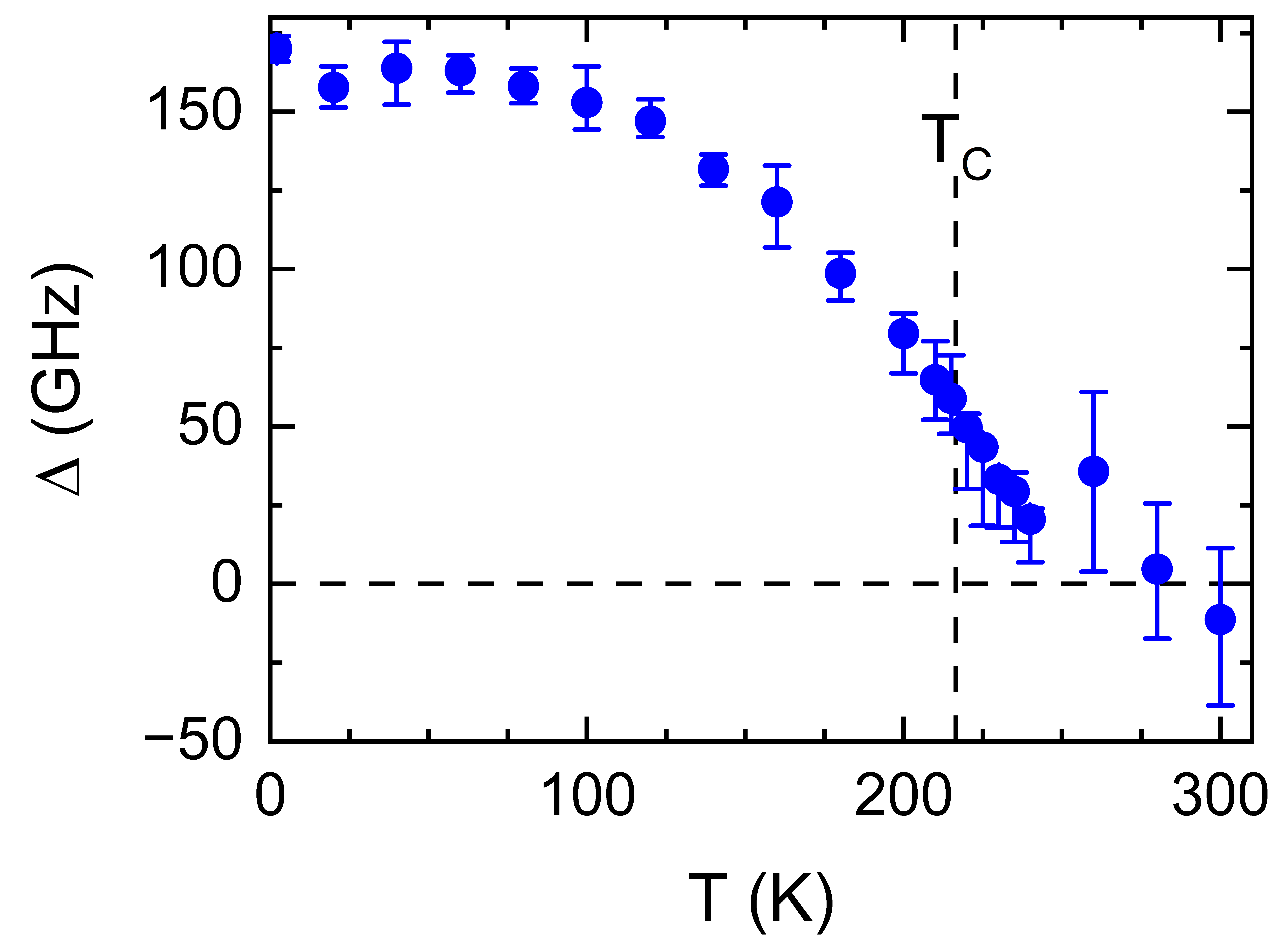}
\caption{Temperature dependence of the anisotropy gap  $\Delta$ as described in the text. The vertical dashed line marks \tc~.}
\label{fig:gap_diff_T}
\end{figure}

\begin{figure}
\centering
\includegraphics[width=\columnwidth,clip]{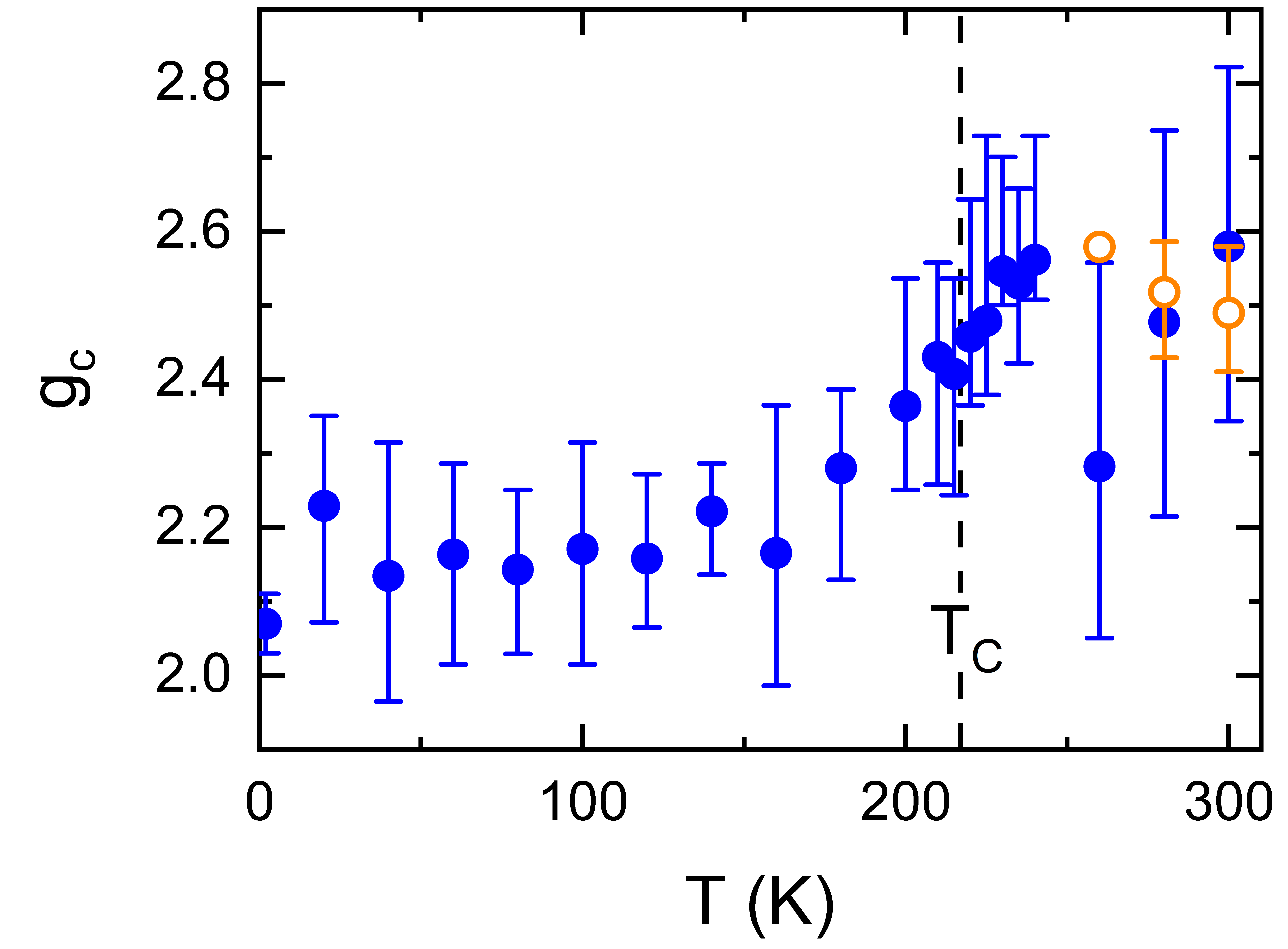}
\caption{Temperature dependence of the effective $g_{\rm c}$-factor determined as described in the text. Open (orange) data points correspond to fits with $\Delta$ fixed to zero. The vertical dashed line marks \tc.}
\label{fig:g_c_diff_T}
\end{figure}

The effect of increasing the temperature on the dynamic magnetic response is studied by means of the temperature evolution of the anisotropy gap and of the effective $g_{\rm c}$ factor. To this end we have recorded $fB$-diagrams for $B||c$ axis for temperatures between $\SI{2}{\kelvin}$ and $\SI{300}{\kelvin}$ (see Fig.~\ref{fig:fb-diagrams_c-axis_diff_T} in the SM~\cite{SM}). For all temperatures, the resonance frequencies linearly depend on the magnetic field. The slope of the resonance branches enables us to determine the effective $g$-factors, $g_{\rm c}$, while their linear extrapolation to zero magnetic field yields the associated excitation gap $\tilde{\Delta}$. The obtained temperature dependencies of the demagnetization-corrected anisotropy gap $\Delta$ and $g_{\rm c}$ are visualized in Fig.~\ref{fig:gap_diff_T} and Fig.~\ref{fig:g_c_diff_T}.

The zero-field anisotropy gap $\Delta$ is rather constant at low temperatures but significantly decreases upon heating (see Fig.~\ref{fig:gap_diff_T}). Notably, $\Delta$ is still finite at \tc~ where it assumes the value of $\approx\SI{65}{\giga\hertz}$, i.e., $\approx\SI{38}{\percent}$ of its low temperature value. We observe a finite gap above \tc\ which only vanishes for $T\gtrsim T_{\Delta}\simeq \SI{270}{\kelvin}$ ($\approx 1.25$\tc). Our data in particular do dot indicate any distinct anomaly at \tc , which implies that on the GHz-timescale the (static) long-range ordering temperature is not decisive. This observation is directly visible in our experimentally observed $fB$ diagrams (see Fig.~\ref{fig:fb-diagrams_c-axis_diff_T} in the SM~\cite{SM}) from which the presence and size of the gap can be directly read off while the demagnetization corrections ($\gamma_{\rm c} N_{\rm z} M_{\rm s}$) are only small. The saturation magnetization $M_{\rm s}$ at each temperature used for the demagnetization correction was approximated by the magnetization measured for $B||c$ at $B = \SI{1}{\tesla} \gg B_{\rm {sat, c}}$ (see Fig.~\ref{fig:magnetization} in the SM~\cite{SM}),  where $B_{\rm {sat,c}}$ denotes the saturation field at \SI{2}{\kelvin} for $B||c$. Consequently, the size of the correction decreases with increasing temperature and amounts to approximately \SI{5}{\giga \hertz} at \tc , which is a magnitude smaller than $\Delta(T=T_{\rm C})$.
    
Concomitantly, the effective $g_c$-factor derived from the data in Fig.~\ref{fig:fb-diagrams_c-axis_diff_T} in the SM~\cite{SM} remains constant at low temperatures, increases above $\sim \SI{160}{\kelvin}$ and again assumes an approximately constant value of $g_{\rm c}\simeq 2.55$ for $T\geq T_{\rm C}$. In order to further reduce the number of fitting parameters, we exploit that, for $T\geq \SI{260}{\kelvin}$, the excitation gap $\tilde \Delta$ assumes values close to $\SI{0}{\giga \hertz}$ (see Fig.~\ref{fig:gap_diff_T}). We conclude that it vanishes in this temperature regime as expected for sufficiently high temperatures. We hence have performed linear fits based on Eq.~\ref{eq:FMR_nu3_domain-based} with $\tilde{\Delta}$ fixed to zero to extract $g_{\rm c}$ above $\SI{260}{\kelvin}$. The resulting $g_{\rm c}$ values exhibit much smaller error bars and suggest that $g_{\rm c}$ slightly decreases with increasing temperature above \tc\ (see Fig.~\ref{fig:g_c_diff_T}). A comparable qualitative behavior of $g_{\rm c}$ with an however more pronounced increase of $g_c$ towards \tc\ was observed in the literature for \cri~\cite{Jonak2022}, \cst~\cite{Li2021} and \cgt~\cite{Li2021, Khan2019}. 

\begin{figure}[htb]
    \centering
    \includegraphics[width=\columnwidth,clip]{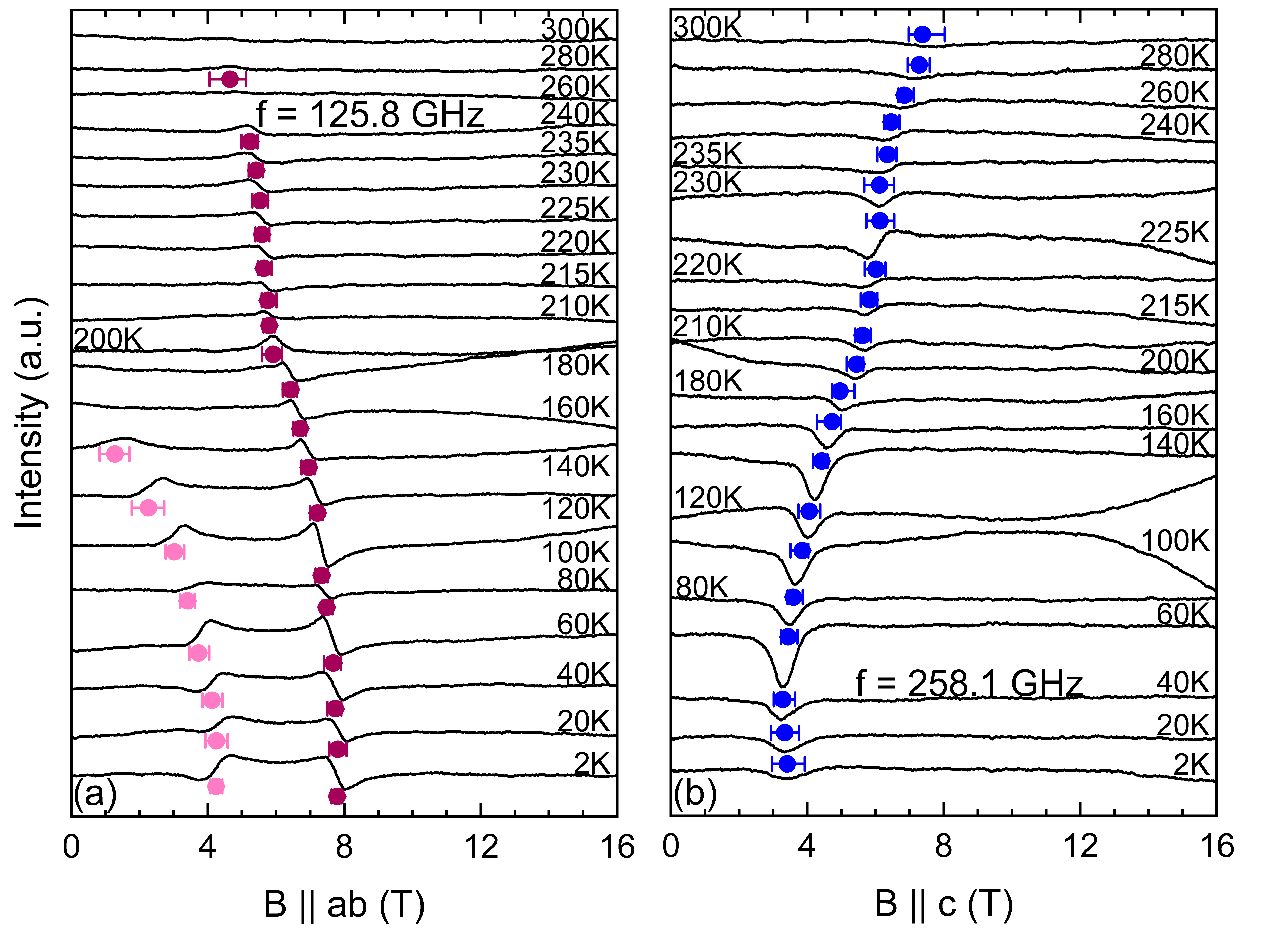}
    \caption{HF-ESR spectra at various temperatures at fixed frequencies for (a) $B||ab$ plane (at $f = $\SI{258.1}{\giga \hertz}) and (b) $B||c$ axis (at $f = $\SI{125.8}{\giga \hertz}). Colored symbols mark the positions of the resonance features. Where needed the spectra were corrected for a linear or quadratic background.}
    \label{fig:spectra_T_dep}
\end{figure}

The observation of a finite anisotropy gap well above \tc\ indicates the presence of quasi-static magnetic order on the time-scale of the experiment, i.e., in the $10^2$-GHz-regime, in this temperature range. To further prove this scenario we studied the evolution of the actual local fields with temperature by measuring the resonance fields at selected frequencies. These measurements have been performed for $B||c$ axis and $B||ab$ plane at $f = $\SI{258.1}{\giga \hertz} and $f =$\SI{125.8}{\giga \hertz}, respectively, in the temperature regime $\SI{2}{\kelvin}\leq T\leq\SI{300}{\kelvin}$. The recorded spectra are shown in Fig.~\ref{fig:spectra_T_dep}. For $B||ab$ plane, we observe two resonance features which both shift to lower magnetic fields with increasing temperature. For $B||c$ axis, in contrast, the resonance shifts to higher magnetic fields upon heating. The shifting of the resonance field positions continues above \tc.

\begin{figure}
	\centering
	\includegraphics[width=\columnwidth,clip]{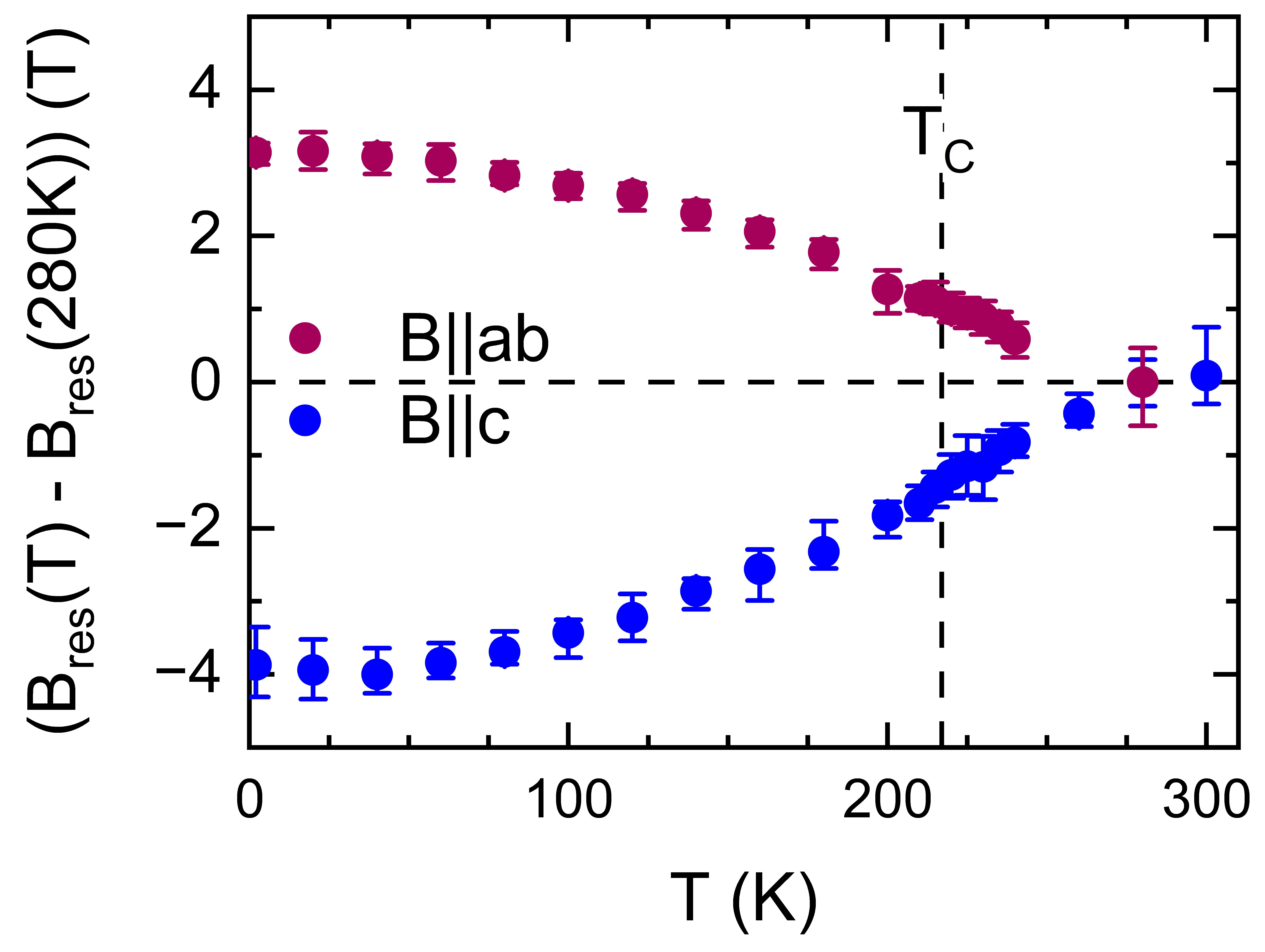}
	\caption{Temperature dependence of the shift of the resonance fields at $f =$\SI{125.8}{\giga \hertz} ($B||ab$ plane) and $f = $\SI{258.1}{\giga \hertz} ($B||c$ axis) from its value at \SI{280}{\kelvin}. The vertical dashed line marks \tc .}
	\label{fig:res_shift_diff_T}
\end{figure}

The resonance fields obtained from the measurements shown in Fig.~\ref{fig:spectra_T_dep} were used to determine the shift of the resonance fields with respect to the resonance field at the highest measured temperature where the resonances for $B||ab$ plane and $B||c$ axis are both still visible ($\SI{280}{K}$; see Fig.~\ref{fig:spectra_T_dep}). This shift is visualized in Fig.~\ref{fig:res_shift_diff_T}. $|B_\mathrm{res}(T)-B_\mathrm{res}(\SI{280}{\kelvin})|$ decreases with increasing temperature for both directions of the external applied magnetic field and vanishes only well above \tc. It should be again noted here that the shift of the resonance position is anisotropic up to $\simeq \SI{270}{\kelvin}$, as the resonance fields for $B||ab$ plane and $B||c$ axis shift in an opposite manner.
    
Since the resonance field positions probe the actual local fields, the observed shifts clearly prove the monotonous increase of quasi-static anisotropic fields upon cooling. Both the temperature dependence of $\Delta$ and of the local fields hence imply anisotropic short-range magnetic order up to at least $\SI{270}{\kelvin} \gg T_{\rm C}$. We again emphasize the absence of any discontinuity at \tc\ which shows that, at the time-scale of our experiments, the evolution of true long-range magnetic order does not yield any significant changes of the local magnetic properties. 

\subsection{Discussion}


\subsubsection{The anisotropy gap $\Delta$ at $T=2$~K}

At $\SI{2}{\kelvin}$, the field dependence of the three observed magnon branches is well captured by the domain-based model as shown in Fig.~\ref{fig:fb-diagram_2K}. The saturation field $B_{\rm {sat}} = B_{\rm A}+N_{\rm y}M_{\rm S} = \SI{5.89\pm0.08}{\tesla}$ for $B||ab$ obtained from the simultaneous fit for $B||ab$ and $B||c$ is in good agreement with the reported magnetization data~\cite{Kim2022}. Our data in particular enable us to precisely determine the anisotropy gap of the magnon branches which, at $T=\SI{2}{\kelvin}$ amounts to $\Delta = \SI{170\pm4}{\giga \hertz}$. The uniaxial anisotropy constant can be calculated from the fit parameters  by $K = M_{\rm S}\Delta/(2\gamma)$, yielding $K = \num{10.50\pm0.23}\times\SI{e6}{\erg\per\cubic\centi\meter}$ at $T = \SI{2}{\kelvin}$. 

\begin{table}[htb]
    \centering
    \begin{tabular}{c|c|c|c|c}
\hline\hline
         & \tc~(K) & $\Delta$ (GHz) & $\Delta$~(meV) & Ref. \\
\hline
Fe\(_{2.92(1)}\)Ge\(_{1.02(3)}\)Te\(_2\) & $216$ & $\num{170\pm4}$ & $\num{0.70\pm0.02}$ & this work\\
Fe\(_{2.86}\)GeTe\(_2\) & $215$ & $\num{230\pm50}^{\rm a}$  & $\num{0.96\pm0.2}$\footnote{Obtained as $\Delta=2KS$ with $K=\SI{0.6\pm0.1}{\milli \electronvolt}$ and a reduced spin $S=\num{0.8\pm0.1}$ as reported in Ref.~\cite{Trainer2022}.} &  \cite{Trainer2022} \\
Fe\(_{2.75}\)GeTe\(_2\) & $150$ & $\num{895\pm50}$ & $\num{3.7\pm0.2}$ & \cite{Calder2019} \\
\hline\hline
\end{tabular}
\caption{Excitation gap in \fgt\ as determined by FMR (this work) and as reported by previous INS studies~\cite{Calder2019,Trainer2022}.}
\label{tab:gaps}
\end{table}

The comparison of the excitation gap determined here with reported data from inelastic neutron scattering (INS) on \fgt\ in Table~\ref{tab:gaps} illustrates that Fe site occupancy is a crucial parameter not only for $\Delta$ but also for \tc . It is supposed to govern \tc\ by altering the density of states and the lattice constants and thereby tuning the magnetic interaction and anisotropy energies~\cite{May2016,Calder2019}. Despite a much larger anisotropy gap which is about five times larger than observed at hand, Fe\(_{2.75}\)GeTe\(_2\) shows significantly reduced ordering temperature of only \SI{150}{\kelvin}~\cite{Calder2019} compared to the two samples with smaller iron deficiency. In a purely 2D limit the relation between \tc , the exchange interaction $J$ and the uniaxial anisotropy constant $K$ -- which is proportional to the anisotropy gap by $K = \Delta M_{\rm S}/(2\gamma)$ -- is given by~\cite{Bander1988,Alahmed2021}

\begin{equation}
T_{\rm C} \propto \frac{J}{\ln(\pi^2 J/K)}.
\label{eq:tc_function_of_JK}
\end{equation}

Applying this estimate, the comparison of the characteristic quantities for the differently Fe-deficient samples in Table~\ref{tab:gaps} implies that magnetic coupling must be significantly reduced in Fe\(_{2.75}\)GeTe\(_2\) as compared to the less deficient materials which corroborates the conclusion of deficiency effects on $both$ $J$ and $\Delta$ in~\cite{May2016}. The anisotropy gap obtained from INS~\cite{Trainer2022} for a sample with a similar iron deficiency as used in this work does not deviate significantly from the one determined in this work by HF-FMR. However, considering the error bars, HF-FMR allows for a much more precise determination of the anisotropy gap as compared to the INS reports~\cite{Trainer2022,Calder2019}.


\subsubsection{Domain structure}

In the literature the domain wall configuration of the $ab$ plane was studied frequently~\cite{Li2018, Leon-Brito2016, Nguyen2018, Fei2018, Yang2022, Yi2017}. Depending on the thickness of the sample, the temperature, the cooling procedure and the applied magnetic field stripe domains, labyrinth domains, wavy-stripes with bubble domains, circular domains with double-domain structure, branching features and bubble domains were observed in the $ab$ plane of~\fgt~\cite{Li2018, Leon-Brito2016, Nguyen2018, Fei2018, Yang2022, Yi2017}. In all of the mentioned studies, where a magnetic field was applied, the field was applied parallel to the $c$ axis. Our FMR experiments in contrast provide information on the domain wall configuration of the sample surface with normal vector perpendicular to the crystallographic $c$ axis and magnetic field applied parallel to the $ab$ plane: The parameter $\alpha = \SI{0\pm11}{\degree}$ present in Eq.~\ref{eq:FMR_nu1_domain-based} denotes the angle between the domain walls and the applied magnetic field direction. In this respect we however note the rather high conductivity of \fgt~\cite{deiseroth2006,Kim2018} implying a penetration depth smaller than about $\SI{10}{\nano\meter}$~\footnote{The penetration depth of the electromagnetic wave into a conductor is given by~\cite{Jackson2014} $\delta \approx \sqrt{\nicefrac{2}{\omega \mu \sigma}}$, where $\omega \approx \SIrange{1e11}{1e12}{\per\second}$ is the angular frequency of the used microwave, $\mu = 1 + \chi$ the permeability of \fgt\ and $\sigma \approx \SIrange{1e5}{1e6}{\ohm\metre}$ is the conductivity of \fgt~\cite{deiseroth2006, Kim2018}. As the magnetic susceptibility $\chi > 0$ for FMs, the largest possible penetration depth is in the order of \SI{10}{\nano \meter}.} so that the applied microwave radiation cannot fully penetrate the sample which exhibits a thickness of $\SI{200}{\micro\meter}$. Our experiment hence probes small portions and edges of the sample and its ferromagnetic domain structure, i.e., the surface of the sample. This is confirmed by observing the HF-FMR signal also when the sample is covered by aluminium foil to exclude transmission through the bulk (see Fig.~\ref{fig:80K_spectrum_wo_alufoil} in the SM~\cite{SM}). For the probed volume, our observation of $\alpha \simeq 0$° indicates that at the sample surface with normal vector perpendicular to the crystallographic $c$ axis the distribution of domain wall orientations is centered parallel to the applied magnetic field direction. This suggests a predominantly stripe domain arrangement at the surface, as illustrated in Fig.~\ref{fig:domain_wall_structure} in the SM~\cite{SM}. 

For the $ab$ plane, it has been shown that the stripe-like domains of the labyrinthine domain structures in \fgt~ for higher temperatures and zero magnetic field tend to align preferentially perpendicular to the sample surface~\cite{Li2018}, which is also the case for the surface measured in the current work at low temperatures and applied magnetic field (see Fig.~\ref{fig:domain_wall_structure} in the SM~\cite{SM}). In~\cri, in contrast to \fgt , the domain structure is characterized by two main orientations of the domain walls: perpendicular and parallel to the applied magnetic field, where the latter is the predominant one~\cite{Jonak2022}.

\begin{table*}[t]
\centering
    \begin{tabular}{l|c|c|c|c|c|c|c}
\hline\hline
         & u/p & \tc~(\unit{\kelvin}) & $T_{\rm \Delta}/T_{\rm C} $ & $\Delta$ (\unit{\giga \hertz}) & $B_A$~(\unit{\tesla}) & $K$~(\unit{\erg \per \cubic \centi \meter})& Ref. \\
\hline

\fgt\ & u & 217 & 1.25 & \num{170\pm4} @\SI{2}{\kelvin} & \num{5.85\pm0.08} @\SI{2}{\kelvin} &$\num{10.50\pm0.23}\times\num{e6}$ @\SI{2}{\kelvin}& this work \\ \hline

Fe\(_{4}\)GeTe\(_2\) & u$^a$
&$\approx \num{270}$& $>1.11$&$\approx 30.6$ ($B||c$) @$\SI{3}{\kelvin}^b$
& $\approx 1.07$ ($B||c$) @ $\SI{3}{\kelvin}^{b}$& $\approx \num{2.9e6}$ ($B||c$) @ $\SI{3}{\kelvin}$ &~\cite{Pal2024} \\
&&&&$\approx 15$ ($B||ab$) @ $\SI{3}{\kelvin}^{b}$ & $\approx \num{0.5}$ ($B||ab$) @ $\SI{3}{\kelvin}^{b}$ &$\approx \num{1.4e6}$ ($B||ab$) @ $\SI{3}{\kelvin}$ &\\ \hline

Fe\(_{5}\)GeTe\(_2\) &p&332&uk&$\approx 6.4$ @ \SI{100}{\kelvin}$^{c}$ 
& $\approx 0.22$ @ \SI{100}{\kelvin}$^{d}$&$\num{4.81e5}$ @ \SI{100}{\kelvin}&~\cite{Alahmed2021} \\ \hline

CrBr\(_3\) &u&37&uk&$\approx 18$ @ \SI{10}{\kelvin}$^d$
&$\approx 0.56$ @ \SI{10}{\kelvin}&$\num{7.2e5}$ @ \SI{1}{\kelvin}&~\cite{Shen2021}\\ \hline

CrI\(_3\) &u&61&$\approx 1.3$&$\num{80\pm1}$ @ \SI{2}{\kelvin}&$\num{2.8\pm0.1}$ @ \SI{2}{\kelvin}&$\num{2.9\pm0.9}\times\num{e6}$ @ \SI{2}{\kelvin}$^e$
&~\cite{Jonak2022}\\

&u&68&uk&$\approx 42$ @ \SI{10}{\kelvin}$^{d}$&$\approx 2.7$ @ \SI{10}{\kelvin}&$\approx \num{2.9e6}$ @ \SI{10}{\kelvin}&~\cite{Shen2021}\\

&u&68&uk&$82.9$ @ \SI{1.5}{\kelvin}$^f$
&$2.86$ @ \SI{1.5}{\kelvin}&$\num{3.1e6}$ @ \SI{1.5}{\kelvin}&~\cite{Dillon1965}\\\hline

CrSiTe$_3$ &u&34
&$\approx 1$&$\approx 33$ @ \SI{10}{\kelvin} $^{d}$&$\approx 1.14$ @ \SI{10}{\kelvin}$^{d,g}$
&$\approx \num{5.4e5}$ @ \SI{10}{\kelvin}$^{d,h}$
&~\cite{Li2022} \\\hline

CrGeTe\(_3\) &&$\approx 68$&$\approx 1$&$\approx 16.5$ @ \SI{10}{\kelvin}$^{d}$&$\approx 0.58$ @ \SI{10}{\kelvin}$^{d,i}$
&&\cite{Li2022} (SM) \\ 

&u&$\num{64.7\pm0.5}$&uk&$\approx 9$ @ \SI{2}{\kelvin}$^{d}$&$\approx 0.3$ @ \SI{2}{\kelvin}$^{d}$&$\num{4.0e5}$ @ \SI{2}{\kelvin}&~\cite{Khan2019} \\

&u&$\num{66\pm1}$&1.5&$\approx 7$ @ \SI{4}{\kelvin}$^{d}$&$\approx 0.25$ @ \SI{4}{\kelvin}$^{d}$&$\num{4.8\pm0.2}\times\num{e5}$ @ \SI{4}{\kelvin}&~\cite{Zeisner2019} \\

&u&$\approx 61$&uk&&$\approx 0.49$ @ \SI{5}{\kelvin}&$\approx \num{3.65e5}$ @ \SI{5}{\kelvin}&~\cite{Zhang2016} \\

&u&67.9&1&$\approx 6$ @ \SI{40}{\kelvin}$^{d}$&$\approx 0.22$ @ \SI{40}{\kelvin}$^{d}$&$\approx \num{1.9e5}$ @ \SI{40}{\kelvin}$^{d,j}$
&~\cite{Wang2023} \\

\hline\hline
\end{tabular}
\footnotesize{\begin{flushleft}$^a$The intrinsic magnetic anisotropy is of easy-axis type~\cite{Pal2024}. A more complex behavior at low temperatures is reported and yield distinct values for $\Delta$ and $B_{\rm A}$ for different magnetic field directions~\cite{Pal2024}. \\

$^b$Obtained as $\Delta = \gamma B_{\rm A} = 2\gamma K/M_{\rm S}$ with $K \approx \SI{2.9e6}{\erg \per \cubic \centi \meter}$ ($B||c$), $K \approx \SI{1.4e6}{\erg \per \cubic \centi \meter}$ ($B||ab$), $g_{\rm c} = \num{2.045\pm0.032}$, $g_{\rm ab} = \num{2.073\pm0.015}$ and $M_{\rm S} \approx \SI{539.19}{\erg \per \cubic \centi \meter \per \gauss}$ from Ref.~\cite{Pal2024}.\\

$^c$Obtained as $\Delta = \gamma B_{\rm A}$ with $B_{\rm A} \approx \SI{0.22}{\tesla}^{d}$, $g \approx 2.1$ from Ref.~\cite{Alahmed2021}.\\

$^d$No demagnetization correction was applied.\\

$^e$Obtained as $K = B_{\rm A} M_{\rm S }/2$ with $B_{\rm A} = \SI{2.81\pm0.1}{\tesla}$ and $M_{\rm S} = \SI{3.0\pm1.0}{\mu_{\rm B} \per \formulaunit}$ from Ref.~\cite{Jonak2022}.\\

$^f$Obtained as $\Delta = \gamma B_{\rm A}$ with $B_{\rm A} = \SI{2.86}{\tesla}$ and $g = 2.07$ from Ref.~\cite{Dillon1965}.\\

$^g$Obtained as $\Delta = \gamma B_{\rm A}$ with $\Delta \approx \SI{33}{\giga\hertz}^{d}$ and $g \approx 2.07$ from Ref.~\cite{Li2022}.\\

$^h$Obtained as $K = B_{\rm A} M_{\rm S}/2$ with $B_{\rm A} \approx \SI{1.14}{\tesla}$, $M_{\rm S} \approx \SI{2.8}{\mu_{\rm B} \per \formulaunit}$~\cite{Li2022} and $\# \mathrm{f.u.}/\mathrm{unit~ cell} = 3$, $V_{\mathrm{unit~cell}} = \SI{0.8301\pm0.0001}{\cubic \pico \meter}$~\cite{Carteaux1995}.\\

$^i$Obtained as $\Delta = \gamma B_{\rm A}$ with $\Delta \approx \SI{16.5}{\giga\hertz}^{d}$ and $g \approx 2.03$ from Ref.~\cite{Li2022} (SM).\\

$^j$Obtained as $K = B_{\rm A} M_{\rm S}/2$ with $B_{\rm A}\approx 0.22^{d}$ and $M_{\rm S} \approx \SI{30}{\erg\per\gauss\per\gram}$~\cite{Wang2023} and $\rho =$\SI{5.68}{\gram\per\cubic\centi\meter}~\cite{MaterialsProject_CrGeTe3}.\\

 \end{flushleft}}

\caption{Microscopic anisotropy-related parameters of FM vdW materials determined by FMR. Listed parameters are: Anisotropy type (uniaxial (u) or planar (p)), critical temperature \tc, relation $\frac{T_{\rm \Delta}}{T_{\rm C}}$ (unknown (uk) revers to samples were FMR was not measured at $T>$\tc), anisotropy gap $\Delta$, anisotropy field $B_{\rm A}$ and anisotropy constant $K$.}
\label{tab:list}
\end{table*}

\subsubsection{Anisotropic short-range order above \tc }

Short-range magnetic correlations in \fgt~  above \tc~ are expected due to its low-dimensional nature but experimental evidence reported in the literature is either ambiguous or appears in micropatterned material: In neutron scattering data reported in Ref.~\cite{Calder2019}, a characteristic ring feature in the $hk$ plane is well visible above \tc\ which might originate from short range magnetic order. However it was stated that, also phonon scattering could lead to this observation~\cite{Calder2019}. Ref.~\cite{Li2018} reported a non-paramagnetic state above \tc\ in micropatterned \fgt\ where magnetic domains are observed up to room temperature~\cite{Li2018}.

The experimental FMR data presented above clearly confirm the presence of short-range magnetic order in \fgt~ well above \tc\ by detecting anisotropic local magnetic fields and a finite anisotropy gap at the \SI{100}{\giga\hertz} timescale. 
From our data, we conclude the existence of three distinct temperature regimes:

(1) For $T \gtrsim T_{\rm \Delta}$ ($T_{\rm \Delta} \sim \SI{270}{\kelvin}$), we find vanishing of $\Delta$ and there are no quasi-static anisotropic magnetic fields. However, at $T \simeq \SI{300}{\kelvin}$ we still observe a rather large $g$ factor of $g_{\rm c} \simeq 2.5$ and different resonance fields for $B||ab$ plane and $B||c$ axis. This suggests that at $\SI{300}{\kelvin}$ a pure PM regime is not yet reached. This agrees to indications for a non-PM state in \fgt~ extending up to room temperature reported in the literature~\cite{Calder2019,Li2018}. (2) For \tc~$\lesssim T \lesssim T_{\rm \Delta}$ we observe clear evidence of anisotropic short range magnetic order, as $\Delta$ is finite and there are anisotropic shifts of the resonance fields. In addition to this, a plateau in $g_{\rm c}$ is obtained in this temperature region. (3) The development of long-range magnetic order in \fgt\ is not associated with clear anomalies in the FMR spectra as we do not observe any anomalies in the temperature evolution of $\Delta$ and $g_{\rm c}$ at \tc. Specifically, while $\Delta$ assumes finite values below $\sim 270$~K and smoothly increases upon cooling, $g_{\rm c}$ starts to decrease below $\sim 230$~K. We conclude that there is a continuous evolution of quasi-static local fields upon cooling. At low temperatures $T \lesssim \SI{80}{\kelvin}$, $g_{\rm c}$, $\Delta$, and the shift of the resonance fields start to saturate which implies rather temperature-independent static local fields.

The evolution of anisotropic quasi-static order and a finite quasi-static anisotropy gap well above \tc\ is at least a rather common feature of 2D ferromagnets. Anisotropic local fields (cf. Fig.~\ref{fig:res_shift_diff_T}) have been reported for the easy-axis vdW ferromagnets \cri~\cite{Jonak2022}, Fe\(_4\)GeTe\(_2\)~\cite{Pal2024} and CrGeTe\(_3\)~\cite{Zeisner2019} as listed in Table~\ref{tab:list}. In addition, a finite anisotropy gap well above \tc\ is also reported for \cri~\cite{Jonak2022}. Notably, its relative size at \tc , i.e., $\Delta(T_{\rm C})/\Delta(T=2~{\rm K})$, is similar in \fgt\ and \cri\ (see Table~\ref{tab:list}). This also holds for and might be associated with the ratio of the characteristic temperatures $T_{\rm \Delta}/T_{\rm C}$. Extending this comparison to the materials where quasi-static anisotropic fields have been detected above \tc\ suggests similar ratios $T_{\rm \Delta}/T_{\rm C}$ in \fgt, \cri~\cite{Jonak2022}, Fe\(_4\)GeTe\(_2\)~\cite{Pal2024} and CrGeTe\(_3\)~\cite{Zeisner2019}~\footnote{For all materials except \cri\ and \fgt\ reported at hand where a finite gap is reported above \tc , $T_{\rm \Delta}$ is determined from the experimental data as the onset temperature of anisotropy in the resonance fields.}. The contrasting behavior $T_{\rm \Delta}\simeq T_{\rm C}$ has been found for CrSiTe$_3$ by by Li~\etal~\cite{Li2022} who also report such a coincidence for \cgt ; for the latter experimental evidence is however  ambiguous since Refs.~\cite{Li2022} and \cite{Wang2023} report $T_{\rm \Delta}\simeq T_{\rm C}$ in contrast to \cite{Zeisner2019} (see Table~\ref{tab:list}). The examples of \fgt\ at hand and \cri\ however clearly demonstrate that in both 2D ferromagnets long-range magnetic order evolves from $anisotropic$ short-range order which extends in a rather large temperature regime above \tc . At $T>T_{\rm \Delta}$ isotropic short-range order is found in \cri~\cite{Jonak2022} which is also very likely to appear in \fgt . Measurements at higher temperatures than feasible with our ESR setup are needed to further confirm this issue.

\section{Summary}

We investigated the low-energy $q=0$ magnon excitations of \fgt~ in external magnetic fields up to 16~T applied along the $c$ axis and within the $ab$ plane. At $\SI{2}{\kelvin}$, the field dependence of the three observed magnon branches is well captured by a semiclassical domain-based model, highlighting the role of magnetic domains in the FMR response of \fgt. Fitting the domain-based model to the experimental data at $\SI{2}{\kelvin}$ yields the microscopic parameters of the system: the anisotropy gap $\Delta = \SI{170\pm4}{\giga\hertz}$ at $B = \SI{0}{\tesla}$, anisotropy field $B_{\rm A} = \SI{5.85\pm0.08}{\tesla}$ and an isotropic effective g-factor $g_{\rm {ab}} = g_{\rm c} = \num{2.07\pm0.04}$, yielding an uniaxial anisotropy constant of $K = \num{10.50\pm0.23}\times\SI{e6}{\erg\per\cubic\centi\meter}$. At \tc , the anisotropy gap remains finite with a value of $\Delta \approx \SI{65}{\giga \hertz}$, corresponding to about $\SI{38}{\percent}$ of its low-temperature value. The gap closes only above $T_{\Delta} \approx \SI{270}{\kelvin}$, i.e., $T_{\Delta}/T_{\rm C}\simeq 1.25$. Notably, the persistence of a finite anisotropy gap above \tc~ is accompanied by anisotropic shifts in the resonance fields, revealing the presence of anisotropic local magnetic fields above \tc. This implies anisotropic short-range magnetic order evidenced by the finite anisotropy gap and the presence of anisotropic shifts in the resonance fields. Our results show that long-range magnetic order in \fgt\ is driven by the magneto-crystalline anisotropy.

\section*{Acknowledgements}

We are grateful for valuable discussions and insightful input by A.~Alfonsov. Support by Deutsche  Forschungsgemeinschaft (DFG) under Germany’s Excellence Strategy EXC2181/1-390900948 (The Heidelberg STRUCTURES Excellence Cluster) is gratefully acknowledged. B.~Beier acknowledges financial support from Studienförderwerk Klaus Murmann of Stiftung der Deutschen Wirtschaft with funds from BMBF (Federal Ministry of Education and Research).


\section{Supplemental Material}

\renewcommand{\thefigure}{S\arabic{figure}}
\renewcommand{\thetable}{S\arabic{table}}
\renewcommand{\theequation}{S\arabic{equation}}
\setcounter{figure}{0}

The Supplemental Materials contains further information on the mounting of the sample and shows characterizing magnetization data as well as further ESR spectra and the $fB$-diagrams for $B||c$ for various temperatures. We also show an alternative fit to the data and provide information on the single-domain model. 

 
\begin{figure}[h]
	\centering
	\includegraphics[width=0.5\columnwidth,clip]{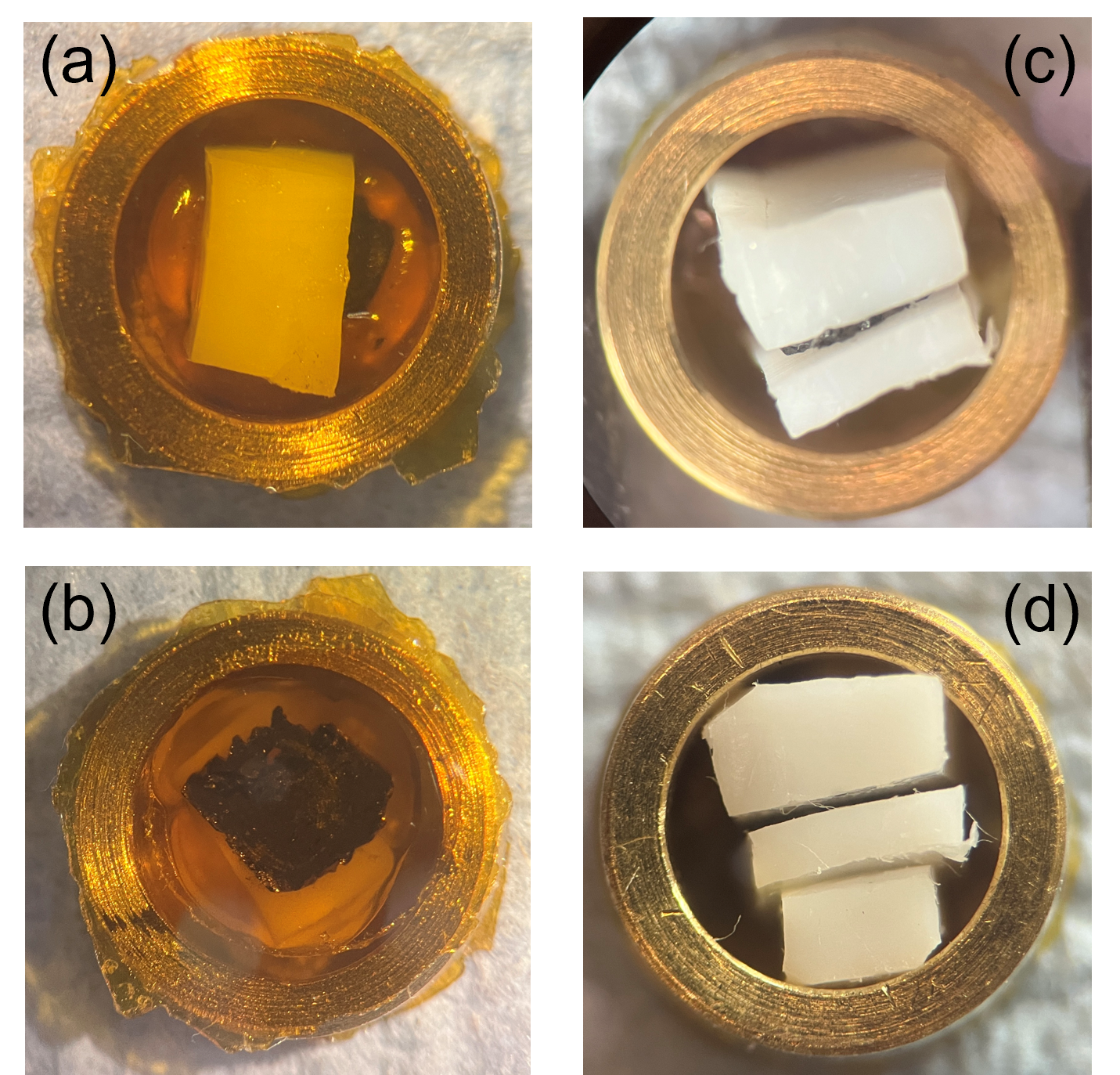}
	\caption{Preparation of the \fgt\ sample in a brass ring sealed on both sides with Kapton tape, where the sticky sides of the Kapton tape are facing each other. (a)/(b) Preparation for $B || c$ axis. The sample is fixed by ApiezonN grease and a teflon block. (c)/(d) Preparation for $B || ab$ plane. The sample is fixed in between two teflon blocks (c) with a minute amount of ApiezonN grease. (d) Stabilisation is ensured by adding a third teflon block and a second brass ring.}
	\label{fig:sample_preparation}
\end{figure}

\begin{figure}[htb]
	\centering
	\includegraphics[width=0.99\columnwidth,clip]{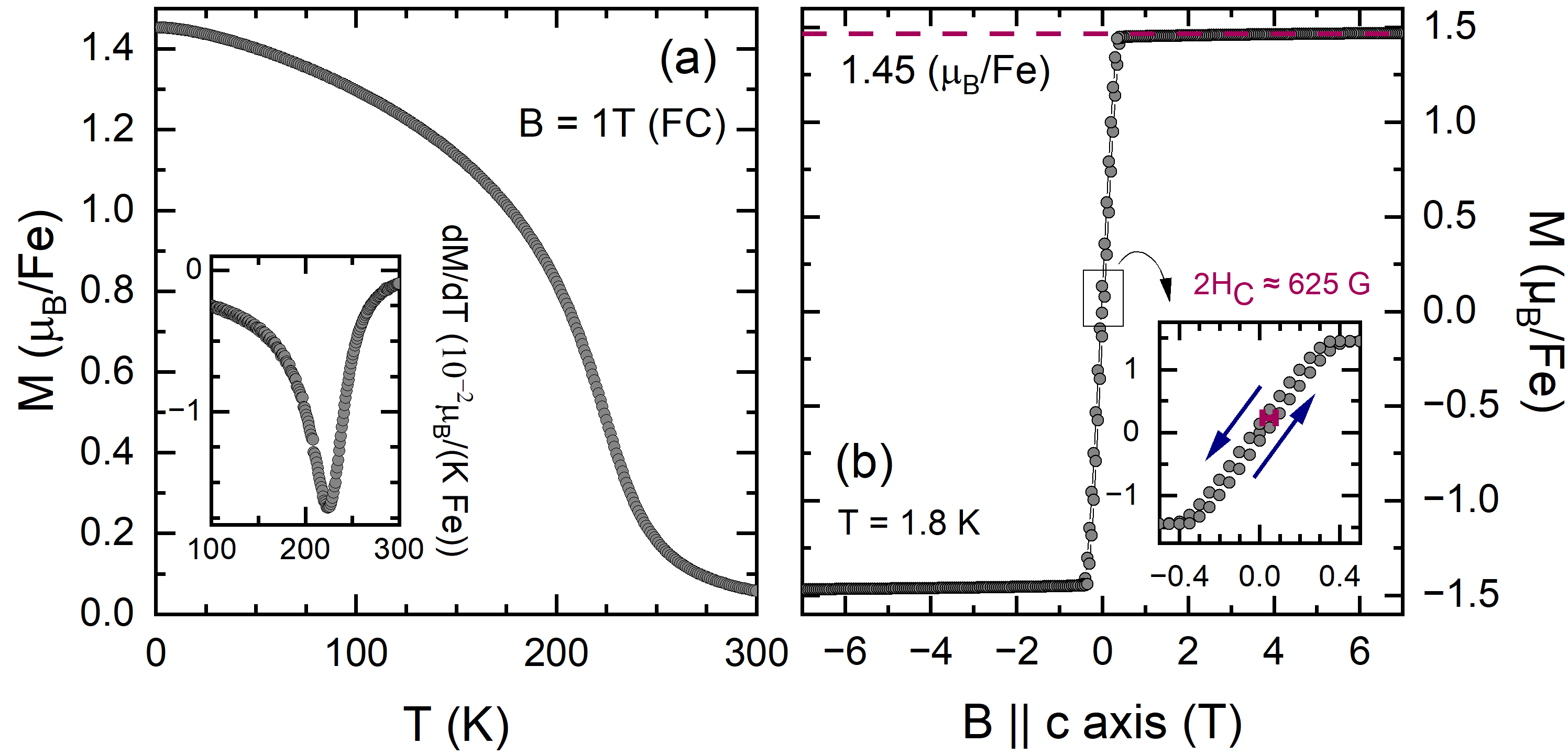}
	\caption{(a) Magnetization as a function of temperature measured field-cooled at $\SI{1}{\tesla}$ with $B || c$ axis. The inset shows the derivative of the magnetization with respect to the temperature. (b) Isothermal magnetization as a function of magnetic field measured at $\SI{1.8}{\kelvin}$ with $B||c$ axis. The horizontal line marks the saturation magnetization. The inset provides a magnified view of the low-field region. Arrows indicate the direction of field sweeps. }
	\label{fig:magnetization}
\end{figure}

\begin{widetext}
\subsection*{Formulas of the field-dependence of the resonance frequency in the single-domain model}

\begin{equation}
\label{eq:FMR_nu1_single-domain}
\left( \frac{\tilde{\nu_1}}{\gamma_{\rm ab}} \right)^2 = \left(\left(B_{\rm A}-\left(N_{\rm z}-N_{\rm y}\right)M_{\rm S}\right)^2-B_{\rm res}^2\right)\frac{B_{\rm A}-\left(N_{\rm y}-N_{\rm x}\right)M_{\rm S}}{B_{\rm A}-\left(N_{\rm z}-N_{\rm y}\right)M_{\rm S}}, \text{for } B\leq B_{\rm A}+(N_{\rm y}-N_{\rm z})M_{\rm S} 
\end{equation}

\begin{equation}
\label{eq:FMR_nu2_single-domain}
\left( \frac{\tilde{\nu_2}}{\gamma_{\rm ab}} \right)^2 = \left(B_{\rm res}-\left(B_{\rm A}-\left(N_{\rm z}-N_{\rm y}\right)M_{\rm S}\right)\right)\\
\cdot\left(B_{\rm res}-\left(N_{\rm y}-N_{\rm x}\right)M_{\rm S}\right),
\text{for } B_{\rm res}\geq B_{\rm A}+(N_{\rm y}-N_{\rm z})M_{\rm S}
\end{equation}

\begin{equation}
\begin{split}
\frac{\tilde{\nu_3}}{\gamma_{\rm c}} = B_{\rm res}+B_{\rm A}-N_{\rm z}M_{\rm S}
\label{eq:FMR_nu3_single-domain}
\end{split}
\end{equation}

\end{widetext}



\begin{figure}[h!]
	\centering
	\includegraphics[width= 0.9\columnwidth,clip]{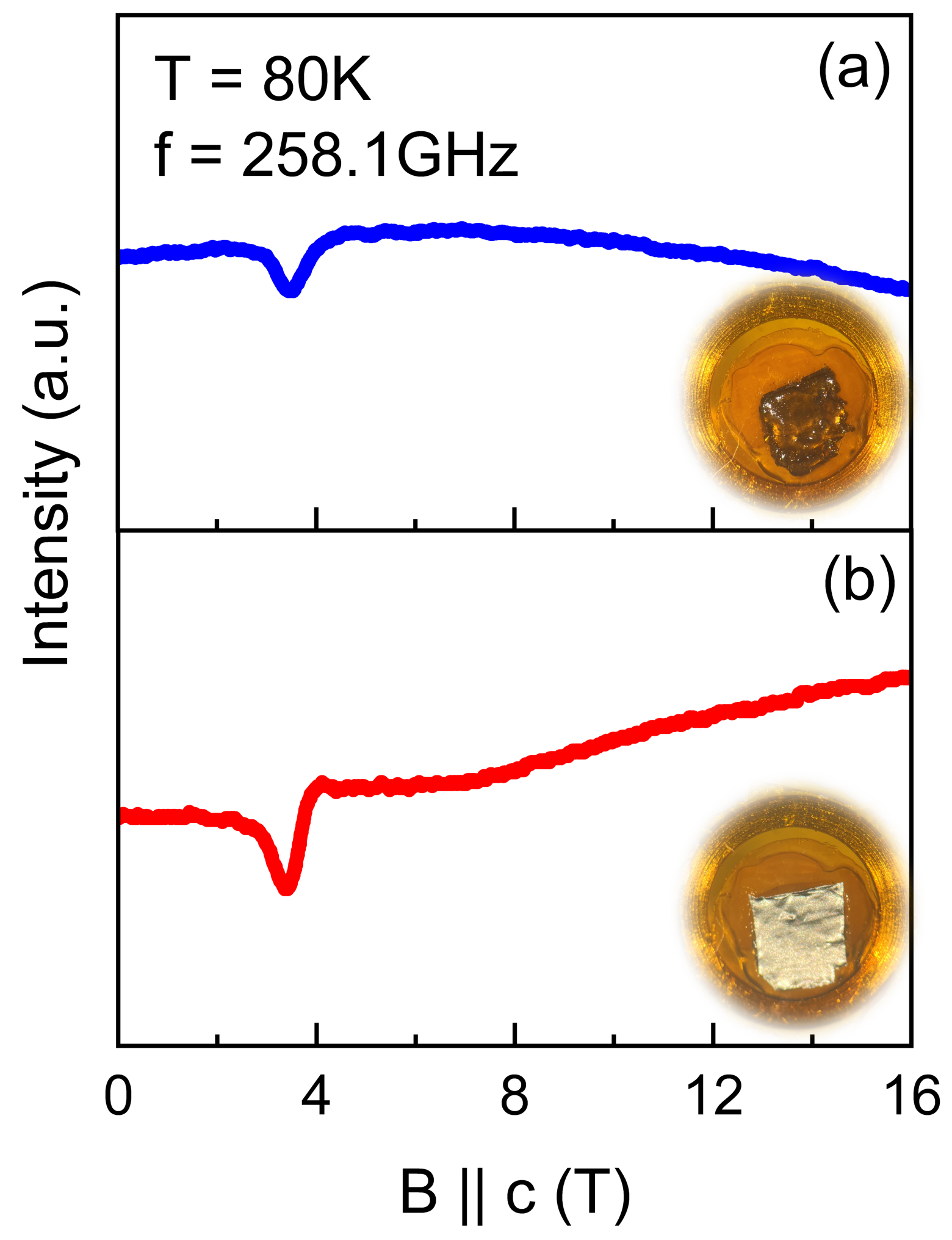}
	\caption{HF-ESR spectra of \fgt\ at \SI{80}{\kelvin} and $f = $\SI{258.1}{\giga \hertz}. In (b) transmission is excluded due to one layer of aluminium foil inbetween the sample and the detection apparatus.}
	\label{fig:80K_spectrum_wo_alufoil}
\end{figure}


\begin{figure}[h!]
	\centering
	\includegraphics[width=\columnwidth,clip]{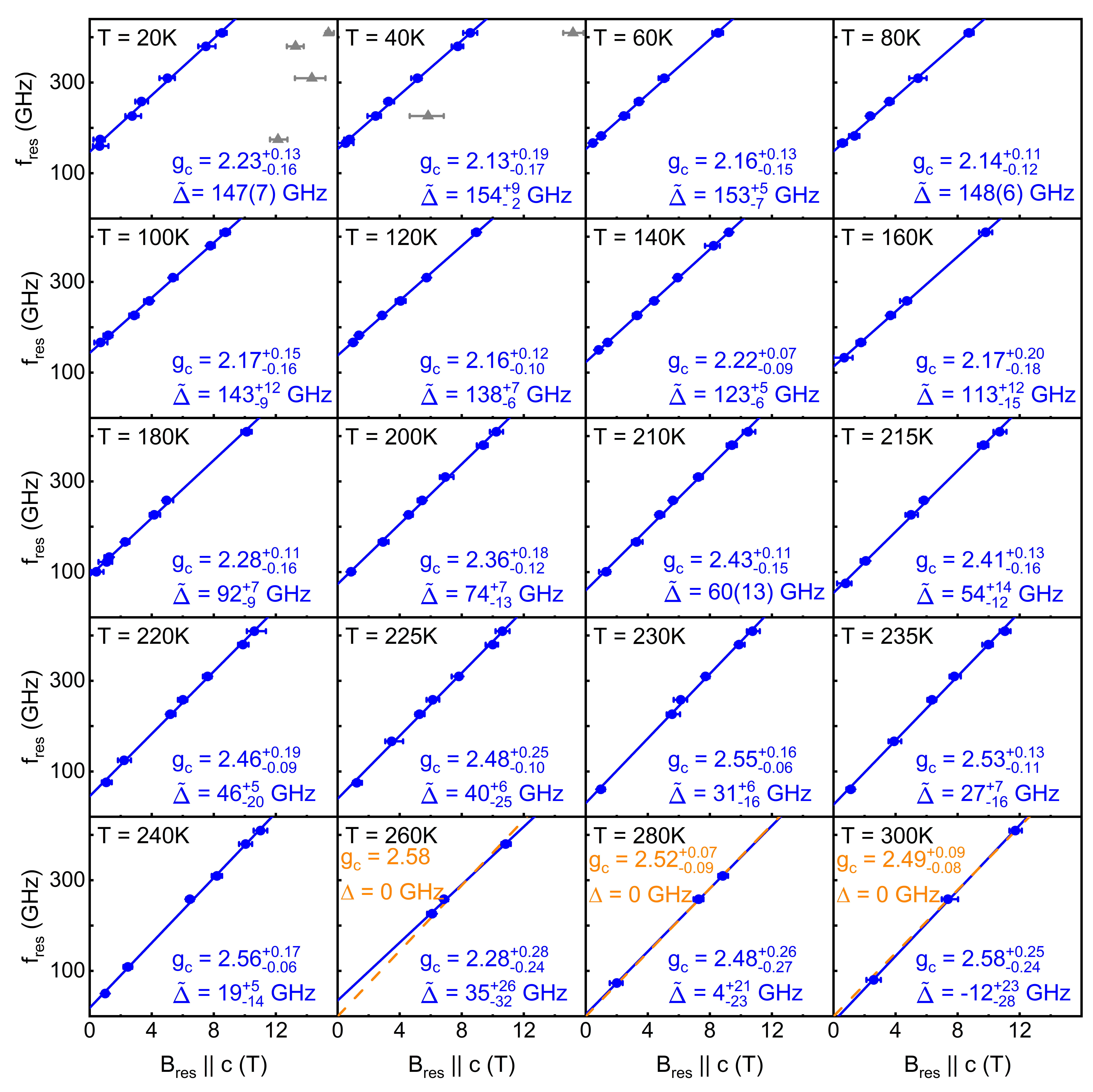}
	\caption{Resonance-frequency--magnetic-field diagrams in \fgt\ for $B||c$ axis at various temperatures. Solid lines are linear fits, the orange lines correspond to linear fits where the excitation gap was fixed to zero. The obtained parameters are displayed in the individual plots.}
	\label{fig:fb-diagrams_c-axis_diff_T}
\end{figure}


\begin{figure}[h!]
	\centering
	\includegraphics[width=0.99\columnwidth,clip]{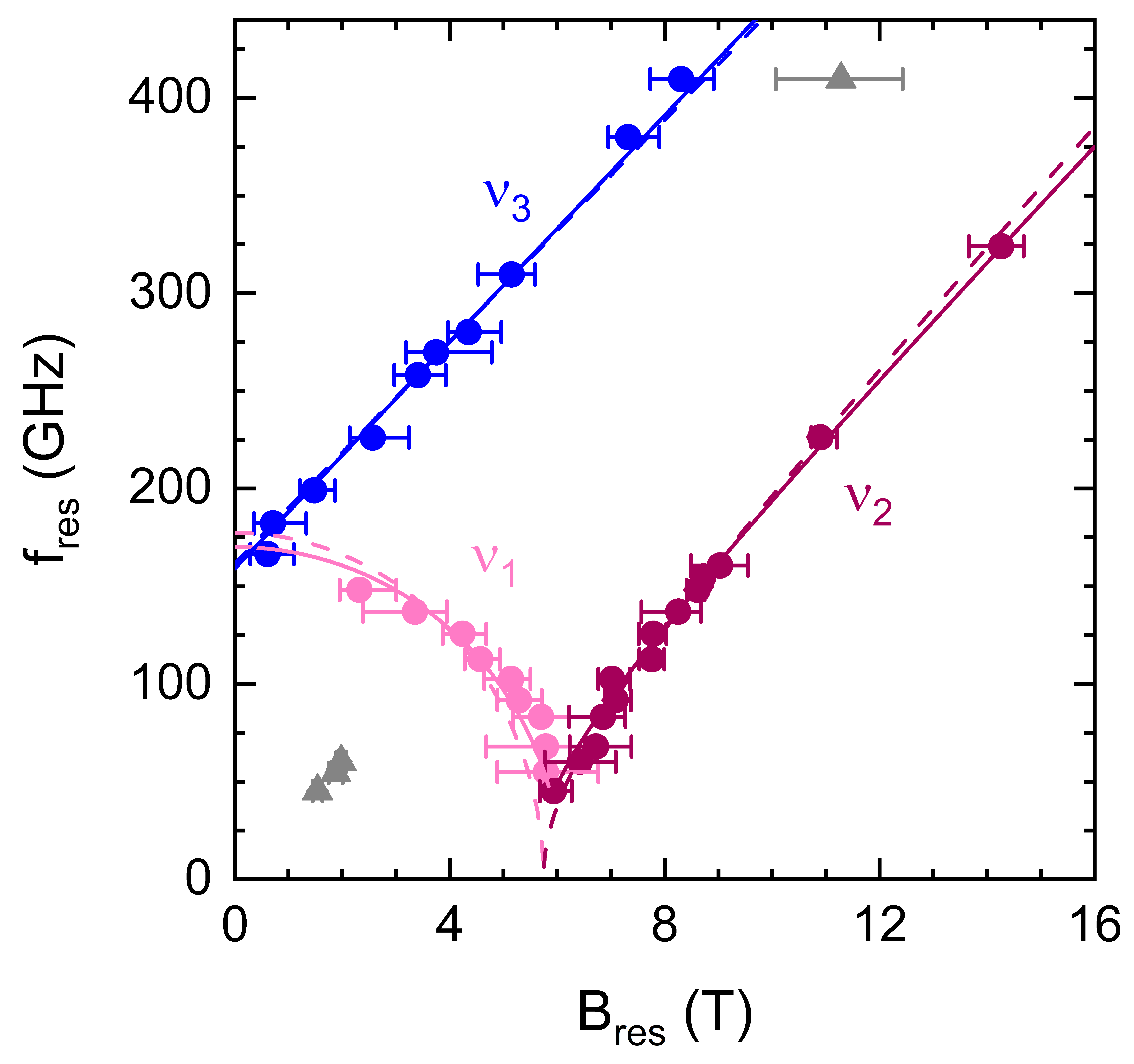}
	\caption{Resonance-frequency-magnetic-field diagram at \SI{2}{\kelvin}. The resonance branches corresponding to the measurement with $B||ab$ plane are labeled $\nu_1$/$\nu_2$ whereas the resonance branch for $B||c$ axis is labeled $\nu_3$. Grey symbols mark other resonances. Solid (dashed) lines are fits to the data using the domain-based model (single-domain model), see text.}
	\label{fig:fb-diagram_2K_supp}
\end{figure}


\begin{figure}[h!]
	\centering
	\includegraphics[width=0.99\columnwidth,clip]{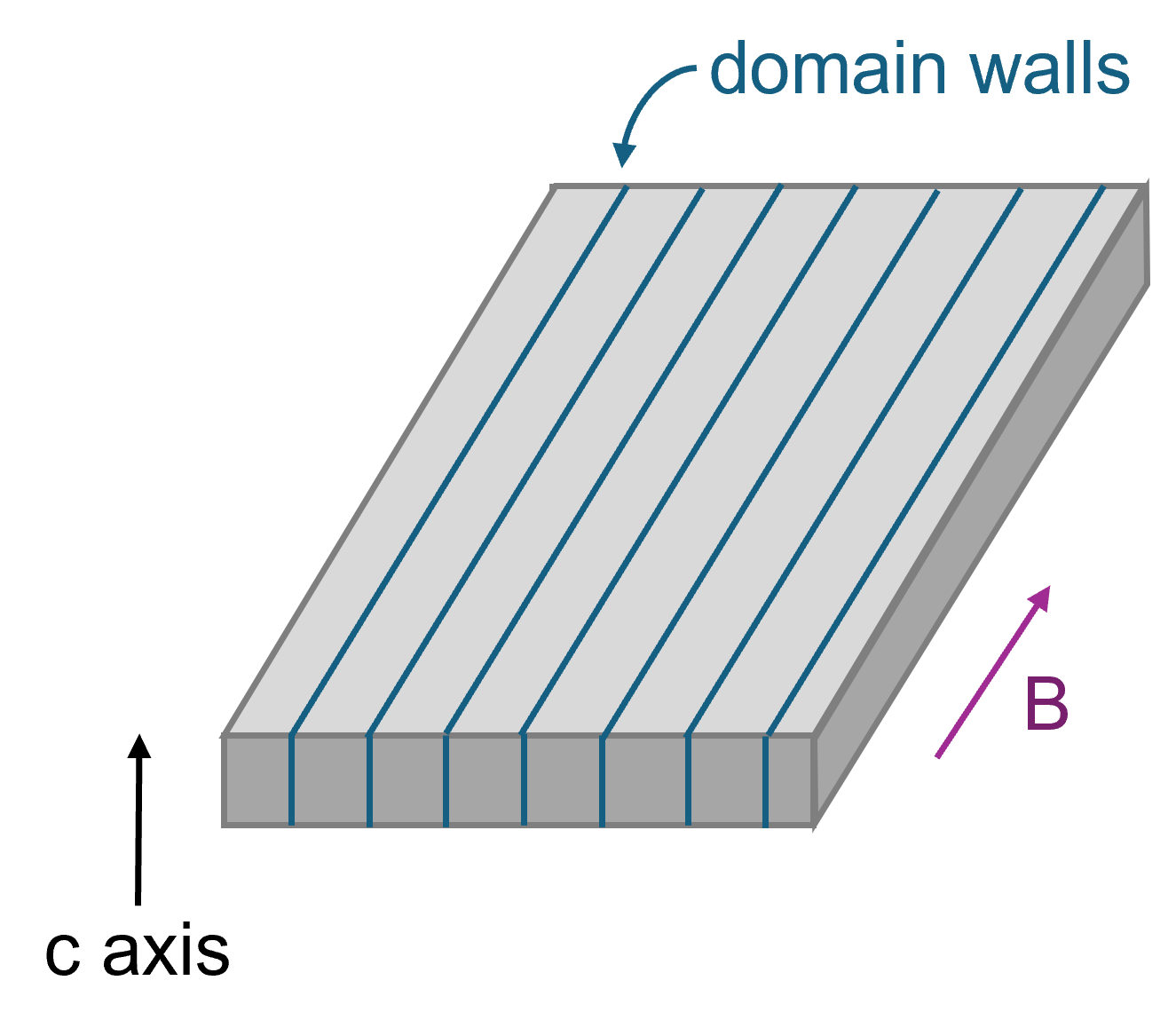}
	\caption{Domain wall structure observed for $B||ab$ plane at the sample surface with normal vector perpendicular to the crystallographic $c$ axis.}
    \label{fig:domain_wall_structure}
\end{figure}

\bibliography{fgt_bibliography}{}
\end{document}